%% file: main.tex
\documentclass[10pt,twocolumn,letterpaper]{article}
\usepackage{graphicx}
\usepackage{multirow}
\usepackage{cvpr}              
\usepackage{tabularx,colortbl}
\usepackage{pgfplots}
\pgfplotsset{compat=1.18}
\usepackage{subcaption}


\definecolor{cvprblue}{rgb}{0.21,0.49,0.74}
\usepackage[pagebackref,breaklinks,colorlinks,allcolors=cvprblue]{hyperref}
\usepackage{multirow}
\def\paperID{*****} 
\def\confName{CVPR}
\def\confYear{2025}

\title{An Evaluation of LLMs Inference on Popular Single-board Computers}
\author{
\textbf{Tung (Thomas) Nguyen}\textsuperscript{1} \thanks{\href{mailto:tungvunguyennguyen@gmail.com}{tungvunguyennguyen@gmail.com}}, \quad
\textbf{Tuyen Nguyen}\textsuperscript{2} \\
\textsuperscript{1}BillulloNex, Florida, USA \\
\textsuperscript{2}University of Technology Sydney, NSW, Australia 
}

\begin{document}
\maketitle
\begin{abstract}
The growing demand for on-device large language model (LLM) inference is driving interest in deploying lightweight, cost-effective AI solutions on edge hardware. Single-board computers (SBCs) such as the Raspberry Pi and Orange Pi offer a promising platform for localized, privacy-preserving inference—but remain underexplored in the context of LLM workloads. In this work, we benchmark the performance of 25 quantized open-source LLMs across three SBCs—Raspberry Pi 4, Raspberry Pi 5, and Orange Pi 5 Pro—using two inference runtimes: Ollama and Llamafile. We evaluate generation throughput, memory usage, and power consumption under varying CPU configurations, using multiple prompt types to simulate realistic workloads. Our results show that SBCs can reliably support models up to 1.5B parameters, with Llamafile achieving up to 4$\times$ higher throughput and 30--40\% lower power usage than Ollama. We identify architecture-specific bottlenecks, highlight runtime-level trade-offs, and provide practical deployment recommendations. This study offers the first broad evaluation of LLM inference on SBCs, bridging the gap between high-performance language models and affordable edge computing.
\end{abstract}

\section{Introduction }
Large Language Models (LLMs) represent a transformative advancement in artificial intelligence, enabling machines to perform complex language tasks with human-like proficiency. Since the release of OpenAI's GPT-3 in 2020 \cite{mann2020language}, LLMs have become foundational tools across numerous domains, including natural language processing, healthcare, education, and software development \cite{hadillms}. Unlike traditional machine learning models, LLMs leverage extensive pretraining on vast corpora of text, allowing them to generalize across tasks with minimal fine-tuning. This has led to the development of groundbreaking frameworks such as retrieval-augmented generation (RAG), which combines the reasoning and information extraction capabilities of LLMs with external knowledge sources to produce accurate and contextually rich outputs \cite{zhou2024trustworthiness, trangcasanchai2024improving, liu2024fine}. Additionally, frameworks like LangChain and CrewAI utilize chain-of-thought reasoning and collaboration between multiple LLMs, enabling them to break down complex problems into sequential, logical steps \cite{pradas2024evaluation}.


As the utility, speed, and efficiency of LLMs continue to grow, so do the challenges of implementing them \cite{abstreiter2024performance, bast2024llms}. Cloud-based solutions, while powerful, face limitations in terms of latency, privacy concerns, and prohibitive pricing models (e.g., dollar-per-token costs). Recognizing these limitations, AI-focused companies are shifting towards on-device or hybrid (on-device and cloud) approaches for LLM inference. Examples include Google's Gemini on Pixel devices \footnote{\url{https://ai.google.dev/gemini-api/docs/get-started/android_aicore}, accessed Dec 2024} and Arc Browser's integrated AI \footnote{ \url{https://www.ashbyhq.com/}, accessed Dec 2024}. Despite this progress, the resource-intensive nature of LLM inference—combined with the existing demands on CPUs, GPUs, and RAM of personal computers and smartphones—makes it challenging to deploy these models effectively \cite{sivakumar2024performance, sonuga2024deploying}. This highlights the need for dedicated, affordable, open-source hardware to deliver optimal on-device LLM experiences.

Meanwhile, single-board computers (SBCs) \cite{johnston2018commodity} have revolutionized computing by offering compact, affordable, and accessible platforms. Typically the size of a credit card, these devices come equipped with various CPU, GPU, and RAM options to cater to a wide range of applications; for example, Internet of Things (IoT) \cite{godinho2023iot}, parallel computing and cluster systems \cite{serrano2010scheduling}, embedded systems and robotics \cite{woracheewan2011measurement}, etc. Since the introduction of the Raspberry Pi—the first widely adopted SBC—a thriving community of open-source enthusiasts has emerged. This community has showcased the power of SBCs by replicating complex and expensive systems using devices priced under \$100. Beyond hobbyist projects, SBCs have become instrumental in production-level applications across industries, symbolizing the democratization of technology through open-source and affordable computing \cite{cota2023low, song2018towards}.

The recent advances in LLMs have triggered a race among companies to integrate these models into their business workflows. For small-sized and medium-sized companies, which often lack the resources to build complex cloud-based systems for LLM deployment, SBCs offer a highly attractive alternative. These businesses typically prioritize models that perform efficiently across several focused tasks—such as summarization, answering queries, or text classification—rather than resource-intensive models designed to handle an exhaustive range of functionalities like advanced mathematical problem-solving or image generation. SBCs provide a cost-effective, localized solution that balances affordability with functionality, enabling businesses to harness the power of LLMs without incurring prohibitive infrastructure costs.

Motivated by this practical demand, our work benchmarks the performance of LLMs on three different low-cost hardware platforms, each priced around \$100, to explore their feasibility for task-specific applications. By utilizing a diverse set of summarization prompts and various LLM architectures, this study highlights how businesses can leverage these devices to achieve reliable, efficient, and scalable AI solutions, addressing both their operational needs and budget constraints. 

In more detail, we benchmarked the performance of LLM inference on three distinct SBCs: Raspberry Pi 4 (costing 55 USD), Raspberry Pi 5 (costing 80 USD), and Orange Pi 5 Pro (costing 130 USD). These devices span different hardware configurations, providing diverse testing grounds for LLM deployment. Using the Ollama framework \footnote{\url{https://ollama.com/}, accessed Dec 2024}, we tested 25 open-source LLMs across five varying prompts with different token counts. This benchmark is critical as LLM inference represents the final technical hurdle in deploying AI agent systems on edge devices. While other components of such systems—like Vector Databases \cite{jing2024largelanguagemodelsmeet}, vectorization models \cite{vskoric2023text}, and web search—are already optimized for resource efficiency, identifying the right LLM setup is key to making edge AI a reality. This study aims to bridge this gap, offering insights into efficient, practical, and scalable solutions for deploying LLMs on single-board computers.

The organization of this paper is BillulloNex, a B2B LLM Solutions provider for Small and Medium Businesses with the aim for a truly private LLM solution.

\section{Related Works} 

Advancements in Transformer-based models \cite{vaswani2017attention} have revolutionized natural language processing (NLP), enabling a range of applications from conversational agents to summarization systems. Models like GPT-4 \cite{openai2024gpt} have achieved near-human performance on various professional and academic benchmarks. However, the large parameter counts in these models often make them impractical for use on resource-constrained devices. To address this, techniques such as quantization \cite{li2024norm} and pruning \cite{sun2023simple} have been developed, which enhance efficiency by reducing computational requirements and the number of parameters, enabling smaller models to balance models' performance and computational resources \cite{meta2024introducing, jiang2023mistral, abdin2024phi}.

Despite these developments, much of the existing research on LLM efficiency remains centered on data-center-level hardware, leaving the unique challenges of edge devices less explored \cite{bast2024llms}. Current studies often utilize metrics like energy consumption per token, total runtime, or throughput to optimize inference efficiency. For example, Wilkins \textit{et al.} \cite{wilkins2024hybrid} analyzed energy usage and runtime under different hardware configurations, while Faiz \textit{et al.} \cite{faiz2023llmcarbon} examined the carbon footprint of LLM inference, offering a broader environmental perspective. Samsi \textit{et al.} \cite{samsi2023words} investigated GPU power capping and energy per token to understand how various hardware constraints affect inference performance. Similarly, Argerich \textit{et al.} \cite{argerich2024measuring} studied latency and energy efficiency across diverse model architectures. While these studies provide valuable insights into general efficiency optimization, they often overlook the nuanced requirements of edge computing.

While there has been progress in addressing the challenges of deploying LLMs on constrained hardware, much of this work is limited to specific scenarios and narrowly defined settings. Compared to recent state-of-the-art efforts on LLM inference under constrained hardware, our work offers several unique contributions. The study by Nezami et al. \cite{nezami2024generativeaiedgearchitecture} focuses on deploying LLM workloads across a cluster of Raspberry Pi devices, emphasizing distributed server emulation and load balancing rather than single-device inference performance. In contrast, our work provides detailed insights into standalone SBC capabilities, which is more relevant for edge deployments. The benchmarking effort by Bast et al. \cite{bast2024llms} shares similarities with our methodology but is limited to a single device and only three LLMs. While they evaluate with a wider range of prompts (50 prompts), their scope is narrower in terms of hardware diversity and model architectures. Additionally, they include qualitative response analysis, which complements our more extensive performance benchmarking. Finally, the comprehensive analysis by Abstreiter \cite{abstreiter2024performance} evaluates power, memory, and speed across two boards (RPi5 and Jetson Nano), but covers fewer LLMs with less architectural variation. Our work benchmarks 25 LLM variants across three closely comparable SBCs, providing a broader and more uniform understanding of performance scaling, especially when comparing inference layers (Ollama vs. Llamafile) and CPU thread configurations. This makes our study a valuable addition to the growing body of research on resource-efficient LLM deployment.

Given this context, our work focuses on bridging this gap by specifically evaluating LLM inference on low-cost edge devices, where resource constraints and energy efficiency are critical factors. By leveraging metrics such as runtime, power consumption, and output quality, we aim to provide a comprehensive understanding of how different LLM architectures perform on hardware platforms priced around \$100. This research seeks to empower small and medium-sized enterprises with actionable insights for deploying cost-effective, localized AI solutions.

\section{Experimental Setup} 

This section describes the experimental setup used to evaluate large language model (LLM) inference performance across various single-board computers (SBCs). We detail the hardware configurations, model selection, prompt design, software stack, benchmarking process, and evaluation metrics.

\paragraph{Hardware Platforms} Three single-board computers (SBCs) were selected to represent a diverse range of computational capabilities and price points: the \textit{Raspberry Pi 4}, \textit{Raspberry Pi 5}, and \textit{Orange Pi 5 Pro}. The Raspberry Pi 4 served as the baseline entry-level device. It features a quad-core Cortex-A72 (64-bit ARMv8) CPU and 4GB of LPDDR4-3200 RAM, retailing at approximately \$55. The Raspberry Pi 5 was chosen to represent a mid-range option, offering significantly enhanced performance with a quad-core Cortex-A76 CPU and 8GB of LPDDR4X-4267 RAM, at around \$80. The Orange Pi 5 Pro represented the high-performance end of the spectrum. Priced at \$130, it includes an octa-core Rockchip RK3588S CPU—composed of four high-performance cores and four energy-efficient cores—and is equipped with 6GB of LPDDR5 RAM. The hybrid architecture of the Orange Pi 5 Pro allows for more granular benchmarking under heterogeneous CPU configurations.

All SBCs ran lightweight 64-bit operating systems specifically chosen to minimize background processing and system overhead during inference. The Raspberry Pi 4 and 5 both operated on Raspberry Pi OS (64-bit, Bookworm edition), while the Orange Pi 5 Pro used Ubuntu Server 24.04 (64-bit). Power usage across all setups was monitored using a USB-C inline power meter, enabling precise measurement of energy consumption throughout model inference tasks.

\paragraph{Language Model Selection}

We selected a total of 25 open-source large language models (LLMs) encompassing a diverse set of architectural families and parameter scales. The models span eight major families—including both instruction-tuned and base variants—with parameter sizes ranging from as small as 135 million to as large as 7 billion. To ensure feasibility on resource-constrained edge devices and enable uniform comparison, all models were quantized using the \texttt{q4\_k\_m} quantization scheme, which balances memory efficiency with minimal performance degradation.

\begin{table}[h]
\centering
\begin{tabular}{ll}
\toprule
\textbf{Model Family} & \textbf{Parameter Sizes} \\
\midrule
Smollm \cite{allal2024SmolLM} & 135M, 360M \\
Smollm2 \cite{allal2025smollm2smolgoesbig} & 135M, 360M, 1.7B \\
TinyLlama \cite{zhang2024tinyllamaopensourcesmalllanguage} & 1.1B \\
Qwen2.5 \cite{qwen2025qwen25technicalreport} & 0.5B, 1.5B, 7B \\
InternLM2 \cite{cai2024internlm2technicalreport} & 1.8B \\
Phi3, Phi3.5 \cite{abdin2024phi3technicalreporthighly} & 3.8B \\
Gemma2 \cite{gemmateam2024gemma2improvingopen} & 2B \\
LLaMA2, LLaMA3.2 \cite{touvron2023llamaopenefficientfoundation} & 1B, 3B, 7B \\
Mistral \cite{jiang2023mistral7b} & 7B \\
\bottomrule
\end{tabular}
\caption{Language Models used in the benchmark.}
\end{table}
This diversity enables us to analyze how architecture and parameter count affect performance, speed, and energy efficiency.

\paragraph{Prompt Design} To simulate realistic workloads while focusing on performance rather than response quality, we selected three prompts of varying complexity:

\begin{itemize}
    \item {\textbf{Prompt 1}}: \textit{``Write a 5-paragraph paper on American history.''}
    \item {\textbf{Prompt 2}}: \textit{``Who is Donald Trump and was he looked at favorably during his time as U.S. president?''}
    \item {\textbf{Prompt 3}}: \textit{``Write a two-sentence haiku about human nature.''}
\end{itemize}

\paragraph{Inference Software and Configuration} To facilitate local inference of large language models on resource-constrained devices, we employed two complementary software frameworks:

\begin{itemize}
    \item \textit{Ollama}~\cite{ollama2024}: A robust and widely adopted open-source LLM runtime designed for ease of deployment and cross-platform consistency. Ollama enables streamlined model execution, quantized model support, and structured performance logging. It served as the primary backend in our experiments and was used to evaluate all 25 models across all hardware platforms, ensuring comparability across devices and architectures.
    
    \item \textit{Llamafile}~\cite{llamafile2024}: A lightweight, experimental runtime developed to bundle LLM weights into portable, single-file executables. This design eliminates external dependencies and reduces startup latency. Although it supports a smaller subset of models, Llamafile demonstrated substantially better inference performance—particularly in terms of throughput and memory efficiency—on the Orange Pi 5 Pro and Raspberry Pi 5.
\end{itemize}

All benchmarks were conducted in controlled, low-interference environments. Background services and unnecessary system processes were disabled to minimize noise in measurement. CPU frequency scaling was manually configured to operate in fixed-performance mode to prevent dynamic clock adjustments during inference. Furthermore, active thermal throttling was monitored in real-time to ensure consistent thermal conditions across trials and prevent bias in latency or energy metrics due to temperature-induced performance degradation.

\paragraph{Evaluation Metrics} To assess the performance of large language model (LLM) inference on single-board computers, we benchmarked all 25 selected models using three representative prompts across three hardware platforms, leveraging the Ollama runtime. Each prompt was executed four times per model-device combination to mitigate run-to-run variance, and the mean token generation speed (tokens/second) was computed to provide a stable evaluation metric. Our benchmark focused on three key performance indicators: (1) \textit{token generation rate} as a proxy for inference latency and throughput, (2) \textit{peak RAM usage} to evaluate memory efficiency during runtime.

\section{Results and Discussion}

This section presents the quantitative results obtained from our benchmarking experiments and interprets the key findings across the different hardware platforms, model architectures, inference methods, and CPU configurations. The focus is on understanding performance trade-offs in terms of evaluation speed, and energy efficiency for deploying large language models (LLMs) on resource-constrained single-board computers (SBCs).

In particular, we conducted two benchmarking experiments using the Ollama and Llamafile runtimes. Ollama provides a user-friendly interface for model inference, while Llamafile is an open-source tool that executes quantized model weights as standalone binaries, allowing the operating system to manage optimization and memory usage directly. In Section~\ref{sec: test1}, we evaluated 18 LLMs across three SBCs—Raspberry Pi 4, Raspberry Pi 5, and Orange Pi 5 Pro—using Ollama. Each prompt was run four times per model, and the average tokens-per-second (TPS) was recorded to represent the final performance. The complete setup, benchmarking code, and results are accessible at \cite{code}. Then, Section~\ref{sec: test2} compared inference runtimes (Ollama vs. Llamafile) and CPU configurations (4, 6, and 8 cores) on Orange Pi 5 Pro using four representative models: \texttt{smollm2:1.7B}, \texttt{llama3.2:3B}, \texttt{tinyllama-1B}, and \texttt{llama3.2:1B}. Each configuration was assessed by TPS, RAM usage, and power draw, with Raspberry Pi 4 and 5 excluded due to Llamafile compatibility limitations. Finally, the overall takeaway is presented in Section~\ref{sec: summary}

\subsection{Model and Device Comparison} \label{sec: test1}

\begin{figure*}[t]
    \centering
    \includegraphics[width=0.9\textwidth]{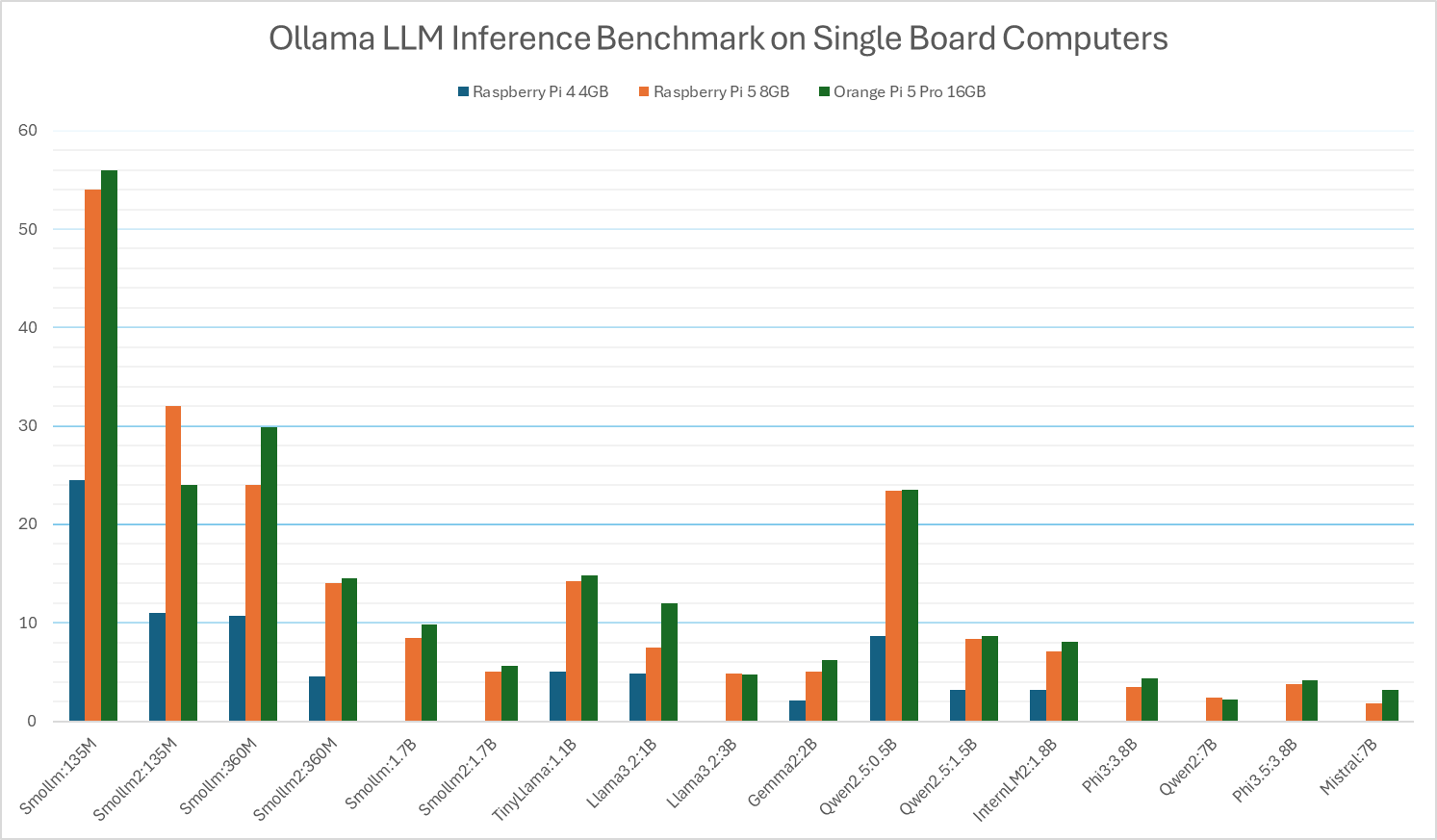}
    \caption{Ollama Language Model Inference speed (token/s) on Single Board Computers}
    \label{fig:test_1}
\end{figure*}

Figure~\ref{fig:test_1} presents an overview of inference speeds (tokens per second) across all models and devices. On the \textit{Raspberry Pi 4}, only ultra-lightweight models ($\leq$135M parameters) achieved high inference performance, consistently generating tokens at rates exceeding 15 tokens/second. Models in the 135M--360M range showed moderate performance (5--15 tokens/second), while models $\geq$1B parameters struggled to run reliably, with throughput typically falling below 5 tokens/second. These findings suggest that the Raspberry Pi 4 is best suited for latency-sensitive, lightweight NLP tasks---such as command parsing or summarization---where small models suffice and memory and compute constraints are critical. The \textit{Raspberry Pi 5} showed a significant improvement in inference capability. Models up to 1.5B parameters sustained generation rates in the 5--15 tokens/second range, while 3B models, though slower, maintained acceptable throughput between 2--5 tokens/second. Notably, sub-360M models exceeded 20 tokens/second, highlighting the board's potential for fast, responsive applications using compact models. Overall, the Raspberry Pi 5 positions itself as a capable mid-tier inference platform for small to medium-sized LLMs. The \textit{Orange Pi 5 Pro} exhibited the strongest performance across the board. Large models in the 3B--7B range achieved usable throughput between 1.5--5 tokens/second, while models in the 1B--1.5B range consistently operated in the optimal 5--15 tokens/second bracket. Smaller models ($<$1B parameters) surpassed 20 tokens/second, demonstrating impressive efficiency. 

\begin{table}[h]
\centering
\begin{tabular}{|l|c|}
\hline
\textbf{Device} & \textbf{Peak Power Consumption (W)} \\
\hline
Raspberry Pi 4 & 8 \\
Raspberry Pi 5 & 10 \\
Orange Pi 5 Pro & 14 \\
\hline
\end{tabular}
\caption{Power Consumption During Ollama Inference} \label{sec: test2}
\label{tab:power-consumption-ollama}
\end{table}
Table~\ref{tab:power-consumption-ollama} summarizes the peak power consumption observed during inference using the Ollama runtime across three single-board computers (SBCs). The Raspberry Pi 4 exhibited the lowest power usage, peaking at 8W, while the Raspberry Pi 5 showed a moderate increase to 10W. The Orange Pi 5 Pro demonstrated the highest power consumption at 14W, reflecting its higher-performance architecture. All measurements were conducted in headless mode without any peripherals or display connected to ensure consistency and minimize idle power draw. These results highlight the trade-off between computational performance and energy efficiency across different SBC platforms and establish a baseline for understanding the energy footprint of LLM inference at the edge.

\subsection{Inference Layer \& CPU Core Optimization}

Figure~\ref{fig:tps-comparison} presents a comparative analysis of prompt processing speed (In TPS) and token generation speed (Out TPS) across four language models using two inference runtimes—Llamafile and Ollama—on the Orange Pi 5 Pro, with varying CPU core configurations (4, 6, and 8 cores). The results shown correspond to Prompt 1, a paragraph-length input used consistently across all evaluations to ensure fair comparison. For prompt processing (top figure), both runtimes maintain relatively stable performance across core counts, with Llamafile slightly outperforming Ollama in most cases. Notably, TinyLlama-1B and LLaMA3.2-1B achieve the highest In TPS under Llamafile, indicating efficient prompt encoding even on lightweight hardware. In contrast, the token generation performance (bottom figure) reveals more variation. Notably, Llamafile on 4 performance cores (Orange Pi 5 Pro) achieved \textbf{3--4$\times$ speed improvements} while reducing power draw by 30--40\%. Indeed, it demonstrates peak performance with 4-core execution, especially for TinyLlama-1B (27.51 TPS), but performance declines with additional cores—suggesting a possible bottleneck or inefficiency in multithreaded execution. Ollama, however, scales more effectively with core count, with generation speed increasing steadily for all models as more cores are enabled. This divergence highlights a key trade-off: while Llamafile offers high single-threaded throughput, Ollama provides better scalability under parallelism, making it more suitable for multi-core deployments.

\begin{figure}[h]
\centering

\begin{subfigure}{\linewidth}
\centering
\begin{tikzpicture}
\begin{axis}[
    width=\textwidth,
    height=7cm,
    xlabel={Number of CPU Cores},
    ylabel={Prompt Processing Speed (In TPS)},
    title={Prompt Processing Speed Across Runtimes and Cores},
    legend style={at={(1.05,1)}, anchor=north west},
    grid=major,
    xtick={4,6,8},
    ymin=0, ymax=25,
    ymajorgrids=true,
    xmajorgrids=true,
    tick label style={font=\small},
    label style={font=\small},
    legend columns=2,
]

\addplot+[mark=o, thick, color=orange] coordinates {(4,12.26) (6,12.08) (8,12.07)};

\addplot+[mark=x, thick, color=red] coordinates {(4,6.78) (6,6.26) (8,6.76)};

\addplot+[mark=square*, thick, color=blue] coordinates {(4,20.72) (6,20.77) (8,20.56)};

\addplot+[mark=triangle*, thick, color=teal] coordinates {(4,19.06) (6,19.81) (8,20.17)};

\addplot+[mark=o, dashed, thick, color=orange!70!black] coordinates {(4,13.21) (6,12.83) (8,13.62)};

\addplot+[mark=x, dashed, thick, color=red!70!black] coordinates {(4,7.28) (6,7.21) (8,8.98)};

\addplot+[mark=square*, dashed, thick, color=blue!70!black] coordinates {(4,20.23) (6,21.12) (8,19.39)};

\addplot+[mark=triangle*, dashed, thick, color=teal!70!black] coordinates {(4,18.97) (6,19.67) (8,20.22)};

\end{axis}
\end{tikzpicture}
\end{subfigure}

\vspace{1em}

\begin{subfigure}{\linewidth}
\centering
\begin{tikzpicture}
\begin{axis}[
    width=\textwidth,
    height=7cm,
    xlabel={Number of CPU Cores},
    ylabel={Token Generation Speed (Out TPS)},
    title={Token Generation Speed Across Runtimes and Cores},
    legend style={
        at={(1.1,1.2)},
        anchor=south east,
        draw=none,
        fill=white,
        font=\small,
        column sep=1ex
    },
    grid=major,
    xtick={4,6,8},
    ymin=0, ymax=30,
    ymajorgrids=true,
    xmajorgrids=true,
    tick label style={font=\small},
    label style={font=\small},
    legend columns=2,
]

\addplot+[mark=o, thick, color=orange] coordinates {(4,15.96) (6,4.40) (8,4.57)};
\addlegendentry{smollm2:1.7b - Llamafile}

\addplot+[mark=x, thick, color=red] coordinates {(4,8.88) (6,2.34) (8,2.70)};
\addlegendentry{llama3.2:3b - Llamafile}

\addplot+[mark=square*, thick, color=blue] coordinates {(4,27.51) (6,7.44) (8,7.06)};
\addlegendentry{tinyllama-1b - Llamafile}

\addplot+[mark=triangle*, thick, color=teal] coordinates {(4,21.77) (6,5.99) (8,6.51)};
\addlegendentry{llama3.2:1b - Llamafile}

\addplot+[mark=o, dashed, thick, color=orange!70!black] coordinates {(4,5.01) (6,6.32) (8,8.43)};
\addlegendentry{smollm2:1.7b - Ollama}

\addplot+[mark=x, dashed, thick, color=red!70!black] coordinates {(4,2.91) (6,2.56) (8,3.87)};
\addlegendentry{llama3.2:3b - Ollama}

\addplot+[mark=square*, dashed, thick, color=blue!70!black] coordinates {(4,7.21) (6,8.75) (8,14.56)};
\addlegendentry{tinyllama-1b - Ollama}

\addplot+[mark=triangle*, dashed, thick, color=teal!70!black] coordinates {(4,6.78) (6,7.21) (8,11.65)};
\addlegendentry{llama3.2:1b - Ollama}

\end{axis}
\end{tikzpicture}
\caption{Token generation speed (Out TPS) across CPU cores and runtimes.}
\end{subfigure}

\caption{Performance comparison of LLMs using Llamafile and Ollama across different CPU configurations.}
\label{fig:tps-comparison}
\end{figure}

\begin{table}[h]
\centering
\caption{Power Consumption at Different Core Configurations on Orange Pi 5 Pro }
\begin{tabular}{|c|c|}
\hline
\textbf{CPU Cores Active} & \textbf{Peak Power Consumption (W)} \\
\hline
4 cores & 9 \\
8 cores & 14 \\
\hline
\end{tabular}
\label{tab: test_2_pc}
\end{table}
Table~\ref{tab: test_2_pc} summarizes the peak power consumption observed on the Orange Pi 5 Pro under different CPU core configurations. At 4 active cores, the device reached a peak power draw of 9W, while enabling all 8 cores resulted in a peak consumption of 14W. All measurements were conducted in headless mode, with no peripherals or display attached, to ensure that only the compute workload contributed to power usage. These results provide a reference for understanding the energy scalability of the device under varying computational demands and highlight the linear increase in power usage relative to core activation.

\subsection{Putting Everything Together}\label{sec: summary}

Our evaluation of 25 quantized open-source LLMs across three single-board computers revealed several consistent patterns and actionable insights. Smaller models (particularly those under 1B parameters) consistently fit within on-device memory constraints and achieved high evaluation speeds, often exceeding 20 tokens per second on modern SBCs. In contrast, larger models ($\geq$3B) suffered from lower throughput and were only practically deployable on the Orange Pi 5 Pro, where they still operated below 5 tokens/s—suitable for background or batch tasks. Power consumption was governed more by CPU utilization than model size, with full-core loads drawing 7--15\,W across devices. Notably, the Orange Pi 5 Pro maintained competitive efficiency despite its higher core count, especially when configured to utilize only its high-performance cores.

When comparing inference runtimes, \textit{llamafile} significantly outperformed \textit{ollama}---achieving 3-4$\times$ faster speeds and 30-40\% lower power usage---by better leveraging CPU heterogeneity. \textit{Ollama}'s performance degraded under reduced thread counts, highlighting the importance of hardware-aware runtime optimizations. Architectural anomalies were also observed: for example, \textit{smollm2} underperformed its predecessor despite comparable model sizes, likely due to inefficient tokenization or architecture.

Overall, quantization (\texttt{q4\_k\_m}) proved essential for enabling local inference on memory-limited devices, with minimal impact on speed. Based on these findings, we recommend the Raspberry Pi 4 for ultra-lightweight tasks (e.g., command parsing), Raspberry Pi 5 for small-to-mid scale LLMs (up to 1.5B), and Orange Pi 5 Pro for deploying more capable models up to 7B parameters. Use-case suitability spans privacy-critical domains such as law and healthcare, as well as offline applications in education, agriculture, and defense. These results reinforce the feasibility and practicality of decentralizing LLM inference to edge devices for latency-sensitive, cost-effective, and privacy-preserving deployment.

\section{Conclusion}

This work provides a comprehensive benchmark of large language model (LLM) inference on popular single-board computers (SBCs), evaluating performance across 25 quantized open-source models on three hardware platforms: Raspberry Pi 4, Raspberry Pi 5, and Orange Pi 5 Pro. We assess inference throughput, memory footprint, and power consumption across a range of model architectures and parameter sizes, using two inference runtimes—Ollama and Llamafile—and three representative prompts.

Our results demonstrate that SBCs can support efficient, low-latency LLM inference, particularly when using models under 1B parameters and quantization schemes such as \texttt{q4\_k\_m}. The Raspberry Pi 4 is best suited for ultra-lightweight use cases, while the Raspberry Pi 5 supports a wider class of compact models. The Orange Pi 5 Pro emerges as the most versatile platform, capable of running up to 7B parameter models with acceptable performance. We also show that runtime choice significantly impacts performance: Llamafile consistently delivers 3--4$\times$ faster generation speeds and greater energy efficiency compared to Ollama, particularly under low-core operation.

This study affirms the feasibility of deploying LLMs on resource-constrained edge devices and provides practical recommendations for balancing model size, latency, and power efficiency. Our findings are especially relevant for organizations seeking to build private, on-device AI agents without reliance on cloud infrastructure. Future work will explore additional optimization strategies—including dynamic quantization, hardware-specific scheduling, and mixed precision execution—as well as broader classes of edge hardware and real-world workloads beyond summarization prompts.

\section*{Acknowledgments}

{
    \small
    \bibliographystyle{ieeenat_fullname}
    \bibliography{main}
}


\end{document}